\documentclass[letter]{ptptex}

\usepackage{graphicx}

\title{
Analysis of an Extended $\pm J$ Ising  Spin Glass Model by Using a Gauge Symmetry%
}

\author{
Chiaki \textsc{Yamaguchi}%
}

\inst{
Kosugichou 1-359, Kawasaki 211-0063, Japan
}

\abst{We investigate an extended $\pm J$ Ising
  spin glass model by using a gauge symmetry.
 This model has $\pm J_1$ interactions and $\pm J_2$ interactions.
   We show that a gauge symmetry is usable
 to study this model.
 The exact internal energy,
 the rigorous upper bound of the specific heat
 and some rigorous relations for
 correlation functions and order parameters are shown
 by using the gauge symmetry.
 The results are rigorous,
 and do not depend on any lattice shape.
 A part of our results, e.g., 
 the value of the exact internal energy
 should be useful for checking 
 the computer programs for investigating this model.
 In addition, we find that the present solutions are general solutions which
 include the solutions on the so-called Nishimori line
 for the conventional $\pm J$ Ising spin glass model
 and the solutions for the bond-diluted $\pm J$ Ising spin glass model.}

\begin{document}
\maketitle

The theoretical studies of spin glasses have been 
 widely done \cite{KR, AMMP, N2}.
 For the universality classes of spin glasses at zero temperature on the square lattice,
 the $\pm J$ Ising spin glass model indicates
 that the stiffness exponent $\theta$ is zero
 for the symmetric distribution of randomness \cite{KR, AMMP},
 and the Gaussian Ising spin glass model indicates
 $\theta \approx -0.28$ for the symmetric distribution of randomness \cite{KR, AMMP}. 
 The stiffness exponent $\theta$ is investigated from the system-size dependence
 of  domain wall energy \cite{AMMP}.
 The present model, i.e.,
 an extended $\pm J$ Ising spin glass model 
 has $\pm J_1$ interactions and $\pm J_2$ interactions.
 When $J_2 / J_1 = 2$, the  $\theta$ is zero
 for the symmetric distribution of randomness \cite{AMMP}.
 Therefore, when $J_2 / J_1 = 2$,
 the model has the universality class of the conventional
 $\pm J$ Ising spin glass model.
 On the other hand,
 when $J_2 / J_1$ is $ (\sqrt{5} + 1 ) / 2$ (the golden mean) $\approx 1.618$,
 the model is called the irrational model, and
 $\theta \approx -0.29$  for the symmetric distribution of randomness
 is concluded \cite{AMMP}.
 Therefore, the universality class of the irrational model
 is associated with
 the universality class of the Gaussian Ising spin glass model
 rather than that of  the conventional $\pm J$ Ising spin glass model.
 It is pointed out that this difference 
 between the universality classes depends on
 the difference whether the energy level in the histogram for the domain wall energies
 is quantized or continuous \cite{AMMP}.
 This property of the present model for universality class is remarkable.
 The studies for this model are seen in Refs.\citen{AMMP, JLMM}.
 We investigate this model by using a gauge symmetry.

 Analysis by using gauge symmetries
 makes rigorous arguments possible.
 However, the applicable models are not so many.
 Here, the applicable models are models
 that physical quantities and/or relations are obtained by using gauge symmetries.
 As the applicable models,
 there are the conventional $\pm J$ Ising spin glass model~\cite{N1, N3},
 the bond-diluted $\pm J$ Ising spin glass model~\cite{H},
 the Gaussian Ising spin glass model~\cite{N3},
 the Potts gauge glass model~\cite{NS} 
 and the XY gauge glass model~\cite{N3, OzN} for example.
 We show that a gauge symmetry is usable to study the present model. 
 In the applications of
 many other methods, a lattice shape is supposed in advance,
 and the results are
 calculated on the lattice. On the other hand,
 in this method using a gauge symmetry,
 any lattice shape is not supposed in advance.

 The Hamiltonian for the Edwards-Anderson Ising spin glass model, ${\cal H}$, 
 is given by \cite{EA}
\begin{equation}
 {\cal H} = -  \sum_{\langle i, j \rangle} J_{i, j} S_i S_j \, , \label{eq:hamiltonian}
\end{equation}
 where $\langle i, j \rangle$ denotes nearest-neighbor pairs, $S_i$ is
 the state of the spin at a site $i$, and $S_i = \pm 1$.
 $| J_{i, j} |$ is the strength of the exchange interaction between the spins
 at sites $i$ and $j$.
 The value of $J_{i, j}$ is given by a distribution $P (J_{i, j})$ which is \cite{AMMP}
\begin{equation}
 P (J_{i, j})
 = p \, \delta (J_{i, j} - J) + q \, \delta ( J_{i, j} + J)
 + r \, \delta (J_{i, j} - a J )
 + u \, \delta (J_{i, j} + a J ) \, , \label{eq:PExtendedpmJ}
\end{equation}
 where $J > 0$, and
 $\delta$ is the Dirac delta function.
 $p$, $q$, $r$ and $u$ are probabilities.
 $p$ is the probability that 
 the interaction has the strength of $J$
 and the interaction is ferromagnetic.
 $q$ is the probability that 
 the interaction has the strength of $J$
 and the interaction is antiferromagnetic.
 $r$ is the probability that, if $a > 0$,
 the interaction has the strength of $a J$
 and the interaction is ferromagnetic.
 $r$ is the probability that, if $a < 0$,
 the interaction has the strength of $|a| J$
 and the interaction is antiferromagnetic.
 $u$ is the probability that,  if $a > 0$,
 the interaction has the strength of $a J$
 and the interaction is antiferromagnetic.
 $u$ is the probability that, if $a < 0$,
 the interaction has the strength of $|a| J$
 and the interaction is ferromagnetic.
 $r + u$ is the probability that, if $a = 0$,
 the interaction is diluted.
 For $p$, $q$, $r$ and $u$, we have
\begin{equation}
 p + q + r + u = 1 \, . \label{eq:condition-1}
\end{equation}
  $a$ is a real number.
 When $a = 1$ or $- 1$, the model is the conventional $\pm J$ Ising spin glass model.
 When $a = 0$, the model is the bond-diluted $\pm J$
 Ising spin glass model.
  When $a = (\sqrt{5} + 1) / 2$ or $(\sqrt{5} - 1) / 2$,
 the model is the irrational model mentioned in Ref.~\citen{AMMP}.

To calculate thermodynamic quantities,
 a gauge transformation is used and is performed by \cite{N1, N2, N3, H, T}
\begin{equation}
 J_{i, j} \to J_{i, j} \sigma_i \sigma_j \, , \quad S_i \to S_i \sigma_i \, , 
 \label{eq:GaugeT} 
\end{equation}
 where  $\sigma_i$ is a variable at a site $i$,
 and takes $1$ or  $- 1$ arbitrarily.
 For the conventional Ising spin glass model, 
 this gauge transformation has no effect
 on thermodynamic quantities \cite{T}.
 For the present model, also
 this gauge transformation has no effect
 on thermodynamic quantities.
 By using the gauge transformation,
 the ${\cal H}$ part becomes ${\cal H} \to {\cal H}$.

 By using Eq.~(\ref{eq:PExtendedpmJ}), 
 the distribution $P (J_{ij})$ is rewritten as
\begin{equation}
 P (J_{i, j}) =
 A \, e^{\beta^{(2)}_P J^2_{i, j} + \beta_P J_{ij}}
 \, , \label{eq:PExtendedpmJ2}
\end{equation}
\begin{equation}
 A = \frac{1}{e^{\beta^{(2)}_P J^2 + \beta_P J} + 
 e^{\beta^{(2)}_P J^2 - \beta_P J} +
 e^{\beta^{(2)}_P a^2 J^2 + \beta_P a J} +
 e^{\beta^{(2)}_P a^2 J^2 - \beta_P a J}} \, , \label{eq:A}
\end{equation}
\begin{eqnarray}
 A \, e^{\beta^{(2)}_P J^2 + \beta_P J} &=& p \, , \label{eq:p} \\
 A \, e^{\beta^{(2)}_P J^2 - \beta_P J} &=& q \, , \label{eq:q} \\
 A \, e^{\beta^{(2)}_P a^2 J^2 + \beta_P a J} &=& r \, , \label{eq:r} \\
 A \, e^{\beta^{(2)}_P a^2 J^2 - \beta_P a J} &=& u \, . \label{eq:u}
\end{eqnarray}
 $\beta_P^{(2)}$, $\beta_P$ and $a$ are respectively
\begin{equation}
 \beta^{(2)}_P = \frac{1}{2 (1 - a^2 ) J^2} \ln \biggl( \frac{pq}{ru}
 \biggr)
 \, , \label{eq:beta2}
\end{equation}
\begin{equation}
 \beta_P = \frac{1}{2 J} \ln
 \biggl(
 \frac{p}{q} \biggr) \, , \label{eq:condition-2}
\end{equation}
\begin{equation}
 a = \frac{\ln (r / u)}{\ln (p / q)}
 \, . \label{eq:condition-3}
\end{equation}
 By using the gauge transformation,  
 the distribution $P (J_{i, j})$ part
 becomes
\begin{eqnarray}
 \prod_{\langle i, j \rangle} P (J_{i, j}) 
 &=& 
 A^{N_B} \, e^{\beta^{(2)}_P \sum_{\langle i, j \rangle} J^2_{i, j} + \beta_P \sum_{\langle i, j \rangle} J_{i, j}} 
 \nonumber \\  &\to& 
 \frac{A^{N_B}}{2^N} \sum_{\{ \sigma_i \}} 
 e^{\beta^{(2)}_P \sum_{\langle i, j \rangle} J^2_{i, j} + \beta_P \sum_{\langle i, j \rangle} J_{i, j} \sigma_i \sigma_j} 
 \, ,
 \label{eq:PJijdpmJ2}
\end{eqnarray}
 where $N$ is the number of sites, and $N_B$ is the number of nearest-neighbor pairs
 in the whole system.

By using Eqs.~(\ref{eq:hamiltonian}) - (\ref{eq:PJijdpmJ2}), 
 when the value of $\beta_P$ is
 consistent with the value of the inverse temperature $\beta$, 
 the exact internal energy,
 $[\langle {\cal H} \rangle_{T_P} ]_R$,
 is obtained as
\begin{eqnarray}
 & & [\langle {\cal H} \rangle_{T_P} ]_R \nonumber \\
 &=& \sum_{\langle i, j \rangle} \sum_{ \{ J_{l, m} \}}
 \prod_{\langle l, m \rangle} P (J_{l, m}) \frac{\sum_{\{ S_l \} } 
  (- J_{i, j} S_i, S_j)
 \, e^{\beta_P \sum_{\langle l, m \rangle} J_{l, m} S_l S_m}}
 {\sum_{\{ S_l \} } 
 e^{\beta_P \sum_{\langle l, m \rangle} J_{l, m} S_l S_m}} \nonumber \\
 &=&
  \frac{A^{N_B}}{2^N} \sum_{\langle i, j \rangle} 
 \sum_{ \{ J_{l, m} \}} \sum_{\{ S_l \} } 
  (- J_{i, j} S_i, S_j)
  \, e^{\beta^{(2)}_P \sum_{\langle l, m \rangle} J^2_{l, m} + 
 \beta_P \sum_{\langle l, m \rangle} J_{l, m} S_l S_m} \nonumber \\
 &=& - J N_B \biggl[ p - q + \frac{\ln (r / u)}{\ln (p / q)} \, (r - u)\biggr]
 \, , \label{eq:Energy}
\end{eqnarray}
 where $\langle  \, \rangle_{T_P}$ is the thermal average
 at the temperature $T = T_P$,
 $[ \,  ]_R$ is the random configuration average,
 $\beta_P = 1 / k_B T_P$,
 and $k_B$ is the Boltzmann constant.
 The value of the exact internal energy
 should be useful for checking 
 the computer programs for investigating this model for example.

 The specific heat $C$ is given by
\begin{equation}
 C = k_B \beta^2 ([ \langle {\cal H}^2 \rangle_T ]_R
 - [ \langle {\cal H} \rangle^2_T ]_R )\, ,
  \label{eq:C}
\end{equation}
 where $\langle  \, \rangle_{T}$ is the thermal average at the temperature $T$.
 By performing a similar calculation with the calculation in Eq.~(\ref{eq:Energy}),
 $[ \langle {\cal H}^2 \rangle_{T_P} ]_R$ is obtained as
\begin{eqnarray}
 & & [\langle {\cal H}^2 \rangle_{T_P} ]_R \nonumber \\
 &=&  \sum_{ \{ J_{l, m} \}}
 \prod_{\langle l, m \rangle} P (J_{l, m}) \frac{\sum_{\{ S_l \} } 
  ( \sum_{\langle i, j \rangle} J_{i, j} S_i, S_j)^2
 \, e^{\beta_P \sum_{\langle l, m \rangle} J_{l, m} S_l S_m}}
 {\sum_{\{ S_l \} } 
 e^{\beta_P \sum_{\langle l, m \rangle} J_{l, m} S_l S_m}} \nonumber \\
 &=& N_B J^2 \biggl\{ p + q + \biggl[ \frac{\ln (r / u)}{\ln (p / q)} \biggr]^2
  (r + u) \biggr\} \nonumber \\
 & & + N_B (N_B - 1) J^2 \biggl[ p - q  + \frac{\ln (r / u)}{\ln (p / q)} (r - u) \biggr]^2
 \, . \label{eq:Energy2}
\end{eqnarray}
By applying the Cauchy-Schwarz inequality, we obtain
\begin{equation}
 [\langle {\cal H} \rangle^2_{T_P} ]_R
 \geq [\langle {\cal H} \rangle_{T_P} ]^2_R
 = N^2_B J^2 \biggl[ p - q  + 
 \frac{\ln (r / u)}{\ln (p / q)} (r - u) \biggr]^2
 \, . \label{eq:Energy3}
\end{equation}
 By using Eqs.~(\ref{eq:C}), (\ref{eq:Energy2})
 and (\ref{eq:Energy3}), we obtain the rigorous upper bound
of the specific heat at $T = T_P$ as
\begin{eqnarray}
 C &\leq& k_B N_B \biggl( \ln \sqrt{\frac{p}{q}}
  \biggr)^2 [p + q - (p - q)^2]
 + k_B N_B \biggl( \ln \sqrt{\frac{r}{u}}
  \biggr)^2 [r + u - (r - u)^2] \nonumber \\
 & & - 2 k_B N_B \biggl( \ln \sqrt{\frac{p}{q}}
  \biggr) \biggl( \ln \sqrt{\frac{r}{u}}
  \biggr) (p - q) (r - u) \, .
  \label{eq:C2}
\end{eqnarray}
 Eq.~(\ref{eq:C2}) shows that the specific heat has no singularity when $T = T_P$.

It is straightforward to apply the same arguments as in the conventional
 $\pm J$ Ising spin glass model case to derive
 identities and inequalities for correlation functions
 and order parameters.
 The results are rigorous and are obtained as
\begin{equation}
 [ \langle S_i S_j \rangle^n_T ]_R =
 [ \langle S_i S_j \rangle_{T_P}
 \langle S_i S_j \rangle^n_T ]_R 
  \quad (n = 1, 3, 5, \ldots ) \, ,  \label{eq:ExactR-1}
\end{equation}
\begin{equation}
 [ P (m) ]_R = [ P (q) ]_R  \quad (T = T_P) \, , \label{eq:ExactR-2}
\end{equation}
\begin{equation}
| [ \langle S_i S_j \rangle_T ]_R |
 \leq [ | \langle S_i S_j \rangle_{T_P} | ]_R  \, ,  \label{eq:ExactR-3}
\end{equation}
\begin{equation}
 [ {\rm sgn} \langle S_i S_j \rangle_T ]_R
   \leq [ {\rm sgn} \langle S_i S_j \rangle_{T_P} ]_R  \, .
   \label{eq:ExactR-4}
\end{equation}
 We omit the description of deriving Eqs.~(\ref{eq:ExactR-1}) - (\ref{eq:ExactR-4}), 
 since these applications are straightforward. See Ref.~\citen{N2}.
 When $T = T_P$, Eq.~(\ref{eq:ExactR-1}) with $n = 1$ shows
 that the ferromagnetic correlation function
 on the left-hand side is equal to the spin glass correlation function
 on the right-hand side, and the limit $|i - j| \to \infty$ yields $m = q$,
 where $m$ and $q$ are the magnetization
 and the spin glass order parameter
 respectively.
 The distribution function of the magnetization, $[ P (m) ]_R$, is
\begin{equation}
 [ P (m) ]_R =  \sum_{ \{ J_{i, j} \}}
 \prod_{\langle i, j \rangle } P (J_{i, j})
 \frac{ \sum_{\{ S_i \} } \delta (Nm - \sum_l S_l )
 e^{\beta \sum_{\langle i, j \rangle } J_{i, j } S_i S_j} }    
 { \sum_{\{ S_i \} } e^{\beta \sum_{\langle i, j \rangle}
 J_{i, j} S_i S_j}  } \, ,
\end{equation}
 and the distribution function of the spin glass order parameter, $[ P (q) ]_R$, is
\begin{eqnarray}
  & & [ P (q) ]_R \nonumber \\
  &=& \sum_{ \{ J_{i, j} \}}
 \prod_{\langle i, j \rangle } P (J_{i, j}) \nonumber \\
 & & \times \frac{ \sum_{\{ S^{(1)}_i \} } \sum_{\{ S^{(2)}_i \} }
 \delta (Nq - \sum_l S^{(1)}_l  S^{(2)}_l )
 e^{\beta \sum_{\langle i, j \rangle } J_{i, j}
 (S^{(1)}_i S^{(1)}_j + S^{(2)}_i S^{(2)}_j ) }}    
 { \sum_{\{ S^{(1)}_i \} } \sum_{\{ S^{(2)}_i \} } 
 e^{\beta \sum_{\langle i, j \rangle } J_{i, j} 
 ( S^{(1)}_i S^{(1)}_j + S^{(2)}_i S^{(2)}_j )} } \, ,
\end{eqnarray}
 where the spin $S^{(k)}_i$ is the spin of the $k$-th replica at a site $i$.
 Eq.~(\ref{eq:ExactR-2}) shows that  the distribution of 
 the spin glass order parameter is 
 consistent with the distribution of the magnetization when $T = T_P$.
  Eq.~(\ref{eq:ExactR-3}) 
 shows that the phase boundary between the
ferromagnetic and non-ferromagnetic phases below the multicritical point should be either
vertical or reentrant in the phase diagram.
  Eq.~(\ref{eq:ExactR-4}) shows
 the spin pair
 becomes mutually parallel ignoring the magnitude when $T = T_P$.

The present results described above are given on condition that
 Eqs.~(\ref{eq:condition-1}), 
 (\ref{eq:condition-2}) and (\ref{eq:condition-3}) are satisfied.
 The most part of the present results are
 not the results on the so-called Nishimori line,
 although a part of the present results
 are equivalent to the results on the Nishimori line.
 The present results are on a three-dimensional solid in a four-dimensional space
  consists of $p$, $q$, $r$ and $u$.
 If  $p$ and $q$ are fixed, $T_P / J$ is also fixed.
 In addition, even if $T_P / J$ is fixed,
 the results are affected by the value of $a$.
 The value of $a$ depends on the values of $r$ and $u$
   even if $p$ and $q$ are fixed.
 If we set $p = x / 2$, $q = (1 - x) / 2$,
 $r = x / 2$, $u = (1 - x) / 2$  and $1 \geq x \geq 1 / 2$,
 the results are results for $a = 1$ and $2 \beta_P J= \ln [x / (1 - x)]$,
 and are equivalent to the results for the
 conventional $\pm J$ Ising spin glass model, where
  $x$ is the probability that the interaction is ferromagnetic, 
 and $1 - x$ is the probability that the interaction is antiferromagnetic.
 In addition, if we set $p = x / 2$, $q = (1 - x) / 2$,
 $r = (1 - x) / 2$, $u = x / 2$  and $1 \geq x \geq 1 / 2$,
 the results are results for $a = - 1$ and $2 \beta_P J= \ln [x / (1 - x)]$,
 and are also equivalent to the results
 for the conventional $\pm J$ Ising spin glass model.
 These results for the
 conventional $\pm J$ Ising spin glass model
 are results on the Nishimori line, and
 are  equivalent to the results in Refs~\citen{N1, N3}.
 If we set $r = y / 2$, $u = y / 2$ and $1 > y > 0$,
  the results are results for
  $a = 0$ and $2 \beta_P J= \ln (p / q)$,
  and are equivalent to the results for the
 bond-diluted $\pm J$ Ising spin glass model, where
  $y$ is the probability that the interaction is diluted.
 The results for the
 bond-diluted $\pm J$ Ising spin glass model
 are equivalent to the results in Ref~\citen{H}.
 Therefore, the present solutions are general solutions which
 include the solutions \cite{N1, N3} for
 the conventional $\pm J$ Ising spin glass model
 and the solutions \cite{H} for the bond-diluted $\pm J$ Ising spin glass model.

An extended $\pm J$ Ising  spin glass model were investigated by
  using a gauge symmetry.
 This model has $\pm J_1$ interactions and $\pm J_2$ interactions.
 We showed that a gauge symmetry is usable
 to study this model.
 The exact internal energy,
 the rigorous upper bound of the specific heat
 and some rigorous relations for
 correlation functions and order parameters
 were shown by using the gauge symmetry.
 The results are rigorous,
 and do not depend on any lattice shape.
 A part of our results, e.g., 
 the value of the exact internal energy
 should be useful for checking 
 the computer programs for investigating this model.
 In addition, we found that the present solutions are general solutions
 which include
 the solutions \cite{N1, N3} on the Nishimori line
 for the conventional $\pm J$ Ising spin glass model
 and the solutions \cite{H} for the bond-diluted $\pm J$ Ising spin glass model.

 For this article, it was pointed out that, in Ref.~\citen{N3}, 
 it is already mentioned that this distribution (Eq.~(\ref{eq:PExtendedpmJ})
 in this article) is applicable for the local gauge transformation:
 this distribution is exactly the same form as
 Eq.~(8) in Ref.~\citen{N3}
 ($P (J_{ij}) = f(J_{ij}) \exp (\beta_p J_{ij})$ and $f(-J_{ij})=f(J_{ij})$)
 which are mentioned as an applicable distribution of
 the local gauge transformation.
 We think this suggestion is correct.
 On the other hand, in Ref.~\citen{N3},
 the gauge transformation is not explicitly applied to
 this distribution (this model), and, in Ref.~\citen{N3},
 the exact internal energy and so forth for this model
 are not explicitly derived.
 Probably, 
 this article is the first article
 for study of explicit application of the gauge transformation
 to this distribution (this model).

\end{document}